\newcommand*{\wn}{cm$^{-1}$}

\def\apjl{Astroph.\ J.\ Lett.\ }
\def\apjss{Astroph.\ J.\ Supp.\ Ser.\ }

\def\cpl{Chem.\ Phys.\ Lett.\ }

\def\jms{J. Mol.\ Spectrosc.\ }

\def\jpc{J. Phys.\ Chem.\ }

\def\josa{J. Opt.\ Soc.\ Am.\ }
\def\josab{J. Opt.\ Soc.\ Am.\ B }

\def\cjp{Can.\ J. Phys.\ }
\def\aa{Astron.\ Astrophys.\  }

\documentclass[aps,pra,twocolumn,superscriptaddress,showpacs]{revtex4}
\usepackage{graphics}
\usepackage{dcolumn}
\usepackage{verbatim}

\begin{document}

\title{
The CO A-X System for Constraining Cosmological Drift \\
of the Proton-Electron Mass Ratio}

\author{E. J. Salumbides}
\affiliation{Department of Physics and Astronomy, and LaserLaB, VU University, De Boelelaan 1081, 1081 HV Amsterdam, The Netherlands}
\affiliation{Department of Physics, University of San Carlos, Cebu City 6000, Philippines}
\author{M. L. Niu}
\affiliation{Department of Physics and Astronomy, and LaserLaB, VU University, De Boelelaan 1081, 1081 HV Amsterdam, The Netherlands}
\author{J. Bagdonaite}
\affiliation{Department of Physics and Astronomy, and LaserLaB, VU University, De Boelelaan 1081, 1081 HV Amsterdam, The Netherlands}
\author{N. de Oliveira}
\affiliation{Synchrotron Soleil, Orme des Merisiers, St Aubin BP 48, 91192, GIF sur Yvette cedex, France}
\author{D. Joyeux}
\affiliation{Synchrotron Soleil, Orme des Merisiers, St Aubin BP 48, 91192, GIF sur Yvette cedex, France}
\author{L. Nahon}
\affiliation{Synchrotron Soleil, Orme des Merisiers, St Aubin BP 48, 91192, GIF sur Yvette cedex, France}
\author{W. Ubachs}
\affiliation{Department of Physics and Astronomy, and LaserLaB, VU University, De Boelelaan 1081, 1081 HV Amsterdam, The Netherlands}

\date{\today}

\begin{abstract}

The $\textrm{A}^1\Pi-\textrm{X}^1\Sigma^+$ band system of carbon monoxide, which has been detected in six highly redshifted galaxies ($z=1.6-2.7$), is identified as a novel probe method to search for possible variations of the proton-electron mass ratio ($\mu$) on cosmological time scales. Laboratory wavelengths of the spectral lines of the A-X ($v$,0) bands for $v=0-9$ have been determined at an accuracy of $\Delta\lambda/\lambda=1.5 \times 10^{-7}$ through VUV Fourier-transform absorption spectroscopy, providing a comprehensive and accurate zero-redshift data set. For the (0,0) and (1,0) bands, two-photon Doppler-free laser spectroscopy has been applied at the $3 \times 10^{-8}$ accuracy level, verifying the absorption data. Sensitivity coefficients $K_{\mu}$ for a varying $\mu$ have been calculated for the CO A-X bands, so that an operational method results to search for $\mu$-variation.

\end{abstract}


\maketitle

\section{Introduction\label{sec:introduction}}

The search for a variation of the dimensionless proton-to-electron mass ratio $\mu=m_p/m_e$ on cosmological time scales can be performed by comparing molecular absorption lines in highly redshifted galaxies with the same lines measured in the laboratory. Through detection of H$_2$ and HD lines in absorbing galaxies  upper limits on $\mu$-variation have been deduced, resulting in a $\Delta\mu/\mu < 1 \times 10^{-5}$ constraint at redshifts $z=2-3.5$, corresponding to look-back times of 10-12 billion years towards the origin of the Universe~\cite{Malec2010,Weerdenburg2011}. For a reliable comparison a database of accurately calibrated laboratory wavelengths must be available and this has been accomplished through laser spectroscopic investigations of the Lyman and Werner band absorption systems in H$_2$~\cite{Philip2007,Salumbides2008} and HD~\cite{Ivanov2008}.
Further, to make such a comparison operational, sensitivity coefficients $K_{\mu}$ must be calculated for all lines in the spectrum; these $K_{\mu}$ represent the wavelength shift induced on a line by a varying $\mu$. Such calculations have been carried out for the H$_2$ molecule using different methods~\cite{Varshalovich1993,Meshkov2006,Ubachs2007} as well as for the HD isotopomer~\cite{Ivanov2010}.

Through radio astronomical observations of the NH$_3$ molecule even tighter constraints at the $\Delta\mu/\mu < 1 \times 10^{-6}$ level could be deduced, because the transitions in the NH$_3$ molecule exhibit much larger $K_{\mu}$ coefficients. However, this ammonia method has only been applied in the two systems where NH$_3$ is detected (B0218+365~\cite{Murphy2008,Kanekar2011} and PKS1830-211~\cite{Henkel2009}) at redshifts $z<1$, or look-back times of 6-7 billion years. In addition the methanol molecule was identified as a molecule with radio frequency transitions exhibiting very large sensitivity coefficients due to the internally hindered rotation mode in the molecule~\cite{Jansen2011,Levshakov2011,Jansen2011b}. Such methanol lines were observed in the single object PKS1830-211~\cite{Muller2011,Ellingsen2012} and a constraint of $\Delta\mu/\mu < 1 \times 10^{-7}$ was derived~\cite{Bagdonaite2012}. Thusfar the radio astronomical observations have been limited to $z<1$.

Because the number of useful H$_2$ high redshift absorber systems is less than ten, additional methods are explored for constraining $\mu$ variation at redshifts $z>1$. Recently a number of high redshift observations were reported on the A$^1\Pi$ - X$^1\Sigma^+$ vacuum ultraviolet absorption system of carbon monoxide, first in Q1439+113 at $z_{abs}=2.42$~\cite{Srianand2008}, then in Q1604+220 at $z=1.64$~\cite{Noterdaeme2009}, in Q1237+064 at $z=2.69$~\cite{Noterdaeme2010}, and finally in three additional systems Q0857+18 at $z=1.73$, Q1047+205 at $z=1.77$ and Q1705+354 at $z=2.04$~\cite{Noterdaeme2011}.
While these observations, all performed with the ESO Very Large Telescope, were mainly used to measure the local cosmic background temperature we propose to use the high resolution spectral observations of CO A-X to search for  $\Delta\mu/\mu$ at these redshifts. We note that, in addition, spectra of the CO A-X system as observed toward Gamma-ray bursts could be used for the same purpose, although the only system with these spectral features detected so far (GRB060807) was observed at a too limited spectral resolution~\cite{Prochaska2009}. The rest-frame wavelengths of the CO A-X bands are in the wavelength range $130-154$ nm, hence longward of Lyman-$\alpha$, so that the CO spectral features in typical quasar spectra will fall outside the region of the Lyman-$\alpha$-forest (provided that the emission redshift of the quasar $z_{em}$ is not too far from the redshift $z_{abs}$ of the intervening galaxy exhibiting the molecular absorption). The occurrence of the Lyman-forest lines is a major obstacle in the search for $\mu$-variation via H$_2$ lines~\cite{Malec2010,Weerdenburg2011}.

The present study provides the ingredients to make the CO A-X system operational for detecting $\mu$ variation from quasar absorption spectra. In order to extract bounds on $\Delta\mu/\mu$ at the competitive level of $< 10^{-5}$ a laboratory wavelength data set at an accuracy of $\Delta\lambda/\lambda = 3 \times 10^{-7}$ is required. The A$^1\Pi$ - X$^1\Sigma^+$ absorption system has been investigated over decades, with an important comprehensive study by Field~\emph{et al.}~\cite{Field1972}, resulting in the most accurate data by Le Floch and coworkers~\cite{Lefloch1987}; the data were compiled for astronomical use by Morton and Noreau~\cite{Morton1994}. The accuracy of those data was however limited to 0.06 \wn\ or $\Delta\lambda/\lambda = 1 \times 10^{-6}$. This is insufficient for an accurate constraint on $\Delta\mu/\mu$, since the laboratory values would make up an essential part of the error budget. Moreover Drabbels~\emph{et al.}~\cite{Drabbels1997} had performed an intricate multi-step excitation study to derive level energies of four (e)-parity levels of A$^1\Pi, v=0, J=1-4$ at an accuracy of 0.002 \wn; these results were found to deviate by -0.04 \wn\ from the classical data~\cite{Lefloch1987,Morton1994}. This has led us to perform two independent studies, a one-photon absorption study employing synchrotron radiation and a two-photon Doppler-free laser-based excitation study, to match the accuracy requirement on the wavelengths for the relevant ($v=0-9,0$) bands of the CO A-X system. In addition a calculation of $K_{\mu}$ sensitivity coefficients for the spectral lines in the CO A-X system was performed.

\section{Fourier transform spectroscopy\label{sec:FTspectroscopy}}

\begin{figure}
\resizebox{0.48\textwidth}{!}{\includegraphics{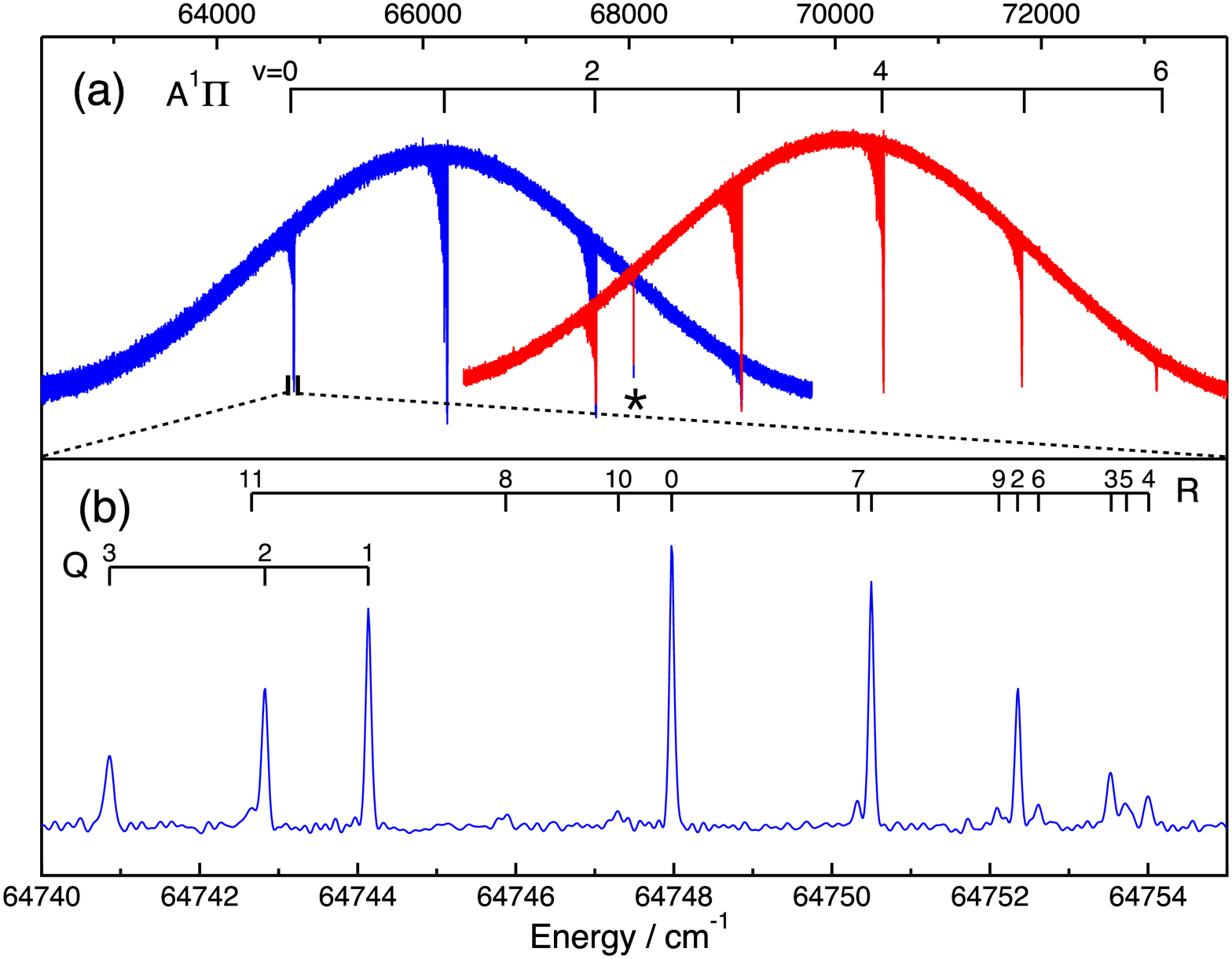}}
\caption{Recording of the jet absorption spectrum of $^{12}$C$^{16}$O with the Fourier-transform vacuum ultraviolet spectrometer at the SOLEIL synchrotron. (a) Overview spectra; (b) Detail spectrum of the A-X (0,0) band. The line indicated by an (*) is a Xe absorption line used for absolute calibration.}
\label{Soleil-spec}
\end{figure}

The unique vacuum-ultraviolet (VUV) Fourier-transform spectrometer (FTS) of the VUV DESIRS beamline~\cite{Nahon2012} at the SOLEIL synchrotron, which is based on wavefront division interferometry~\cite{Oliveira2011}, was employed to record high-resolution absorption spectra of CO in the range $130-160$ nm. The setup was optimized for obtaining spectra at the highest resolution: a molecular slit jet was used to reduce the Doppler width, and the FTS was set to its lowest linewidth of 0.075~\wn, corresponding to a resolving power of about $900\,000$. In Fig.~\ref{Soleil-spec} an overview spectrum, as well as a detail spectrum  is presented showing sharp lines in the A$^1\Pi$-X$^1\Sigma^+$ (0,0) band at a linewidth of 0.09~\wn. Wavelength calibration of the FTS spectrum is derived from the scan-controlling He-Ne laser~\cite{Oliveira2011} and further optimized for the absolute scale referencing to a Xenon line at $68\,045.156$ (3) \wn~\cite{Saloman2004}, which is based on accurate relative measurements~\cite{Humphreys1970} and the measurement of a VUV anchor line~\cite{Brandi2001}. Based on these calibration procedures the uncertainty in the line positions of most CO resonances is estimated to be within 0.01~\wn, corresponding to $\Delta\lambda/\lambda = 1.5 \times 10^{-7}$. Resulting transition frequencies for the CO A-X ($v,0$) bands are listed in Table~\ref{Table-CO-lines-results}. When comparing the results from this high-resolution VUV-FTS study with the classical spectral data~\cite{Lefloch1987,Morton1994} we find an offset of -0.03 \wn\ between the data sets. These offsets are found to be in agreement with those of Drabbels~\emph{et al.}~\cite{Drabbels1997} for the four levels studied.

\begin{table*}
\caption{Calibrated 1-photon transition frequencies (in vacuum \wn) in the A$^1\Pi$ -X$^1\Sigma^+$ ($v$,0) bands. Lines indicated with (*) are derived from the Doppler-free 2-photon laser excitation study of the (0,0) and (1,0) bands and ground state level energies~\cite{Varberg1992}. The other lines are derived from the VUV-FTS study.
\label{Table-CO-lines-results}}
\begin{ruledtabular}
\begin{tabular}{cdddc@{\hspace{20pt}}ddd}
$J''$ & \multicolumn{1}{c}{R} & \multicolumn{1}{c}{Q} & \multicolumn{1}{c}{P} & &\multicolumn{1}{c}{R} & \multicolumn{1}{c}{Q} & \multicolumn{1}{c}{P} \\[1ex]
\colrule
\\[-1.5ex]
	 &				\multicolumn{3}{c}{\textbf{(0,0)}}	&	&				\multicolumn{3}{c}{\textbf{(1,0)}}	\\
	0&	64747.983\,(2)*	&			&			&	&66236.228\,(2)*	&			&			\\	
	1&	64750.504\,(2)*	&	64744.140\,(2)*	&			&	&66238.502\,(2)*	&	66232.384\,(2)*	&			\\
	2&	64752.359\,(2)*	&	64742.828\,(2)*	&	64736.448\,(2)*	&	&66240.016\,(2)*	&	66230.813\,(2)*	&	66224.694\,(2)*	 \\
	3&	64753.54\,(1)	&	64740.863\,(2)*	&	64731.279\,(2)*	&	&66240.798\,(2)*	&	66228.482\,(2)*	&	66219.278\,(2)*	 \\
	4&	64754.01\,(1)	&	64738.233\,(2)*	&	64725.446\,(2)*	&	&66240.85\,(1)	&	66225.422\,(2)*	&	66213.103\,(2)*	\\
	5&	64753.73\,(1)	&	64734.934\,(2)*	&	64718.92\,(1)	&	&66240.25\,(1)	&	66221.654\,(2)*	&	66206.198\,(2)*	\\
\\[-1.5ex]
	 &				\multicolumn{3}{c}{\textbf{(2,0)}}	&	&			\multicolumn{3}{c}{\textbf{(3,0)}}\\
	0&	67678.89\,(1)&				&			&	&69091.62\,(1)	&			&			\\
	1&	67681.27\,(1)&		67675.04\,(1)	&                       &	&69093.90\,(1)	&	69087.78\,(1)   &                       \\
	2&	67682.93\,(1)&		67673.59\,(1)	&	67667.35\,(1)	&	&69095.42\,(1)	&	69086.23\,(1)	&	69080.09\,(1)	\\
	3&	67683.85\,(1)&		67671.39\,(1)	&	67662.05\,(1)	&	&69096.17\,(1)	&	69083.90\,(1)	&	69074.70\,(1)	\\
	4&	67684.06\,(1)&		67668.47\,(1)	&	67656.01\,(1)	&	&69096.12\,(1)	&	69080.79\,(1)	&	69068.52\,(1)	\\
	5&	67683.51\,(1)&		67664.82\,(1)	&	67649.25\,(1)	&	&69095.31\,(1)	&	69076.90\,(1)	&	69061.57\,(1)	\\
\\[-1.5ex]
	 &				\multicolumn{3}{c}{\textbf{(4,0)}}	&	&			\multicolumn{3}{c}{\textbf{(5,0)}}	\\
	0&	70469.92\,(1)&				&			&	&71811.97\,(1)	&			&	   		\\
	1&	70472.07\,(1)&		70466.10\,(1)	&                       &	&71814.05\,(1)	&	71808.12\,(1)   &       		\\
	2&	70473.41\,(1)&		70464.43\,(1)	&	70458.45\,(1)   &	&71815.28\,(1)	&	71806.38\,(1)	&	71800.42\,(1)	\\	
	3&	70473.85\,(1)&		70461.92\,(1)	&	70452.84\,(1)   &	&71815.64\,(1)	&	71803.76\,(1)	&	71794.84\,(1)	\\	
	4&	70473.49\,(1)&		70458.56\,(1)	&	70446.46\,(1)   &	&71815.16\,(1)	&	71800.29\,(1)	&	71788.39\,(1)	\\	
	5&	70472.34\,(1)&		70454.35\,(1)	&	70439.25\,(1)   &	&71813.78\,(1)	&	71795.94\,(1)	&	71781.08\,(1)	\\	
\\[-1.5ex]
	 &				\multicolumn{3}{c}{\textbf{(6,0)}}	&	&			\multicolumn{3}{c}{\textbf{(7,0)}}	\\
	0&	73119.52\,(1)&				&			&	&74394.47\,(1)	&			&			\\
	1&	73121.52\,(1)&		73115.67\,(1)	&                       &       &74396.38\,(1)	&	74390.62\,(1)	&                       \\
	2&	73122.57\,(1)&		73113.83\,(1)	&	73108.00\,(1)	&	&74397.33\,(1)	&	74388.68\,(1)	&	74382.95\,(1)	\\
	3&	73122.67\,(1)&		73111.04\,(1)	&	73102.29\,(1)	&	&74397.30\,(1)	&	74385.79\,(1)	&	74377.16\,(1)	\\
	4&	73121.83\,(1)&		73107.29\,(1)	&	73095.65\,(1)	&	&74396.33\,(1)	&	74381.92\,(1)	&	74370.40\,(1)	\\
	5&	73119.97\,(1)&		73102.60\,(1)	&	73088.08\,(1)	&	&74394.38\,(1)	&	74377.10\,(1)	&	74362.71\,(1)	\\
\\[-1.5ex]
	 &				\multicolumn{3}{c}{\textbf{(8,0)}}	&	&			\multicolumn{3}{c}{\textbf{(9,0)}}	\\
	0&	75634.35\,(1)&				&			&	&76840.23\,(2)	&  			&			\\
	1&	75636.15\,(1)&		75630.52\,(1)	&	                &       &76841.92\,(2)	&	76836.43\,(2)   &                       \\
	2&	75636.94\,(1)&		75628.46\,(1)	&	75622.80\,(1)	&	&76842.60\,(2)	&	76834.26\,(2)	&	76828.72\,(2)	\\
	3&	75636.73\,(1)&		75625.42\,(1)	&	75616.92\,(1)	&	&76842.16\,(2)	&	76831.03\,(2)	&	76822.77\,(2)	\\
	4&	75635.46\,(1)&		75621.33\,(1)	&	75610.03\,(1)	&	&76840.71\,(2)	&	76826.78\,(2)	&	76815.61\,(2)	\\
	5&	75633.20\,(1)&		75616.24\,(1)	&	75602.12\,(1)	&	&76838.14\,(2)	&	76821.50\,(2)	&	76807.63\,(2)	\\
\end{tabular}
\end{ruledtabular}
\end{table*}

\section{UV laser spectroscopy\label{sec:UVspectroscopy}}

\begin{figure}
\resizebox{0.48\textwidth}{!}{\includegraphics{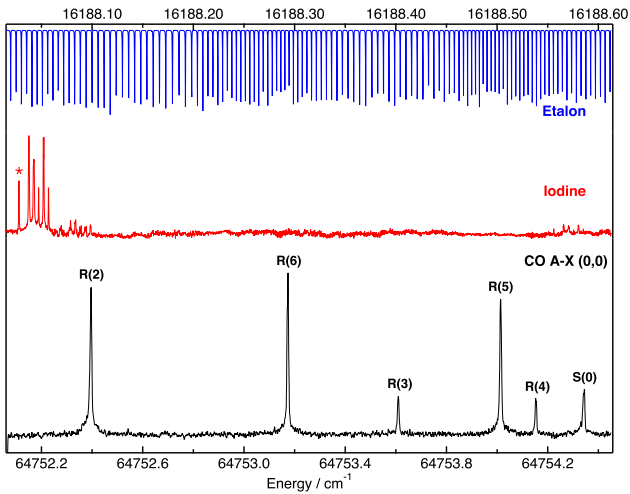}}
\caption{Recording of part of the two-photon laser excitation spectrum of the A$^1\Pi$ - X$^1\Sigma^+$ (0,0) band
of CO. The etalon markers and the I$_2$ hyperfine line (*) recorded in the fundamental are used for frequency calibration.}
\label{CO-PDA-laser}
\end{figure}

In addition two-photon excitation studies on the CO A$^1\Pi$ -X$^1\Sigma^+$ system were performed employing a pulsed dye amplifier (PDA)-laser, injection-seeded by the output of a continuous-wave ring dye laser, delivering narrowband output at $\sim$ 300 nm~\cite{Ubachs1997}. Simultaneous recording of I$_2$ hyperfine lines~\cite{Xu2000,Bodermann2002} and transmission peaks from a stabilized etalon using the fundamental wavelength provides frequency calibration. The excitation was performed in a counter-propagating beam geometry to suppress Doppler effects~\cite{Hannemann2006}. In a 2+1' resonance enhance multi-photon ionization (REMPI) scheme a second laser at 207 nm is employed for ionizing the A$^1\Pi$ excited state population; this laser is pulse delayed by $\sim 10$ ns with respect to the first laser pulse to avoid AC-Stark effects by the ionizing laser pulse. Delayed pulsed-electric-field ion extraction is also employed to minimize DC-Stark effects. Figure~\ref{CO-PDA-laser} shows a recording of an excitation spectrum of part of the A-X (0,0) band. Due to availability and wavelength tunability of the ring dye laser the laser studies were performed on the A-X (1,0) and (0,0) bands.

The error budget in Table~\ref{PDA-uncertainty} lists the estimate of uncertainty contributions from various sources.
AC-Stark shift corrections were obtained from intensity-dependent measurements and extrapolating the transition frequencies to zero intensity levels. The accuracy of the obtained transition frequencies from the laser measurements are limited by the frequency chirp in the PDA-laser. Based on previous characterization of the chirp of our PDA system~\cite{Ubachs1997}, we estimate a 60 MHz contribution from the chirp for this study since a lower-order frequency upconversion of the laser was used. Furthermore, the measurements were performed at lower PDA-pump energies compared to Ref.~\cite{Ubachs1997}, which also reduce the chirp effect. We estimate the uncertainty of the line frequencies from the PDA investigation to be 70~MHz or 0.002~\wn.

In view of parity selection rules in the two-photon laser experiment opposite parity $\Lambda$-doublet components in the A$^1\Pi$ state are excited, when compared to the one-photon absorption experiment. Accurate values for the A$^1\Pi$ $\Lambda$-doublet splittings can be deduced from measurements in the S, Q, R and P branches (the O branch was not recorded), and from the accurately known ground state level energies~\cite{Varberg1992}. Based on this analysis the laser data were converted to transition frequencies for one-photon A-X bands at an accuracy of 0.002~\wn, and listed in Table~\ref{Table-CO-lines-results}.

The results from the laser-based study serve a two-fold purpose. First, they provide a reliable database for comparison with quasar data at the level of $\Delta\lambda/\lambda = 3 \times 10^{-8}$ for the A-X (0,0) and (1,0) bands. Second, a comparison between the present laser data and the VUV-FTS data show agreement between both data sets at a statistical uncertainty within 0.01 \wn. Hence, we conclude that the calibration procedures on the VUV-FTS spectra yield transition frequencies within the estimated uncertainty of 0.01 \wn. This provides a database of CO A-X lines for bands ($v=2-8$) at $\Delta\lambda/\lambda = 1.5 \times 10^{-7}$ accuracy. A slightly worse uncertainty estimate of 0.02 \wn\ is quoted for the (9,0) band which has worse
SNR compared to the other bands due to its lower transition strengths.

\begin{table}
\caption{\label{PDA-uncertainty}
Estimated uncertainties (in MHz) for the transition frequencies measured with the PDA-laser system.
}
\begin{ruledtabular}
\begin{tabular}{lr}
\textrm{Source}&
\textrm{Uncertainty (MHz)}\\
\colrule
Line-fitting		&2\\
Residual Doppler	&$<1$\\
I$_2$ calibration	&2\\
etalon nonlinearity	&2\\
AC-Stark shift		&20\\
DC-Stark shift		&$<1$\\
PDA chirp		&60\\
\colrule
Statistical		&30\\
\colrule
Total			&70\\
\end{tabular}
\end{ruledtabular}
\end{table}

\section{Sensitivity coefficients\label{sec:Kcoefficients}}

For extracting a possible variation of the proton-to-electron mass ratio $\mu$ a set of highly redshifted wavelengths $\lambda_z^i$ are compared with a set of rest-frame wavelengths $\lambda_0^i$ via~\cite{Ubachs2007}:
\begin{equation}
 \frac{\lambda_z^i}{\lambda_0^i}=(1+z_{abs})(1+K^i_{\mu} \frac{\Delta\mu}{\mu})
 \label{Eq-Wavelength}
\end{equation}
where $z_{abs}$ is the redshift of the intervening galaxy absorbing CO  and $K^i_{\mu}$ the sensitivity coefficient for each line to a variation of $\mu$:
\begin{equation}
 K^i_{\mu} = \frac{d \ln \lambda^i}{d \ln \mu} = -\frac{\mu}{E^i_e - E^i_g} (\frac{dE^i_e}{d\mu}- \frac{dE^i_g}{d\mu})
 \label{eq-K}
\end{equation}
with $E^i_e$ and $E^i_g$ the excited and ground state energies connecting an optical transition in the molecule.
It is noted that the definition of Eq.~(\ref{eq-K}) yields $K_{\mu}$ coefficients of opposite sign with respect to other studies, where the definition $\Delta\nu/\nu=K_{\mu}\Delta\mu/\mu$ is used (see e.g. Ref.~\cite{Jansen2011,Jansen2011b}). The specific definition of Eq.~(\ref{eq-K}) is consistent with its use in Eq.~(\ref{Eq-Wavelength}). The relative variation of the proton-electron mass ratio is defined as $\Delta\mu \equiv \mu_z -\mu_0$, meaning that a positive value for $\Delta\mu/\mu$ indicates a larger value of $\mu$ in the cosmological past.

Rovibrational level energies $E(v,J)$ can be expressed in terms of a Dunham expansion
\begin{equation}
 E(v,J) = \sum\limits_{k,l} Y_{k,l} \left(v + \frac{1}{2} \right)^k [J(J+1)- \Lambda^2]^l
 \label{Eq-energy}
\end{equation}
with $Y_{k,l}$ the Dunham coefficients, known to sufficient accuracy for the X$^1\Sigma^+$ state~\cite{Guelachvili1983} and for the A$^1\Pi$ state~\cite{Simmons1969}. $\Lambda$ represents the electronic angular momentum, $\Lambda = 0$ for X$^1\Sigma^+$ and $\Lambda = 1$ for A$^1\Pi$. The advantage of the Dunham representation of molecular states is that the coefficients scale as $Y_{k,l} \propto \mu^{-l-k/2}_{red}$, with $\mu_{red}$ the reduced mass of the molecule. In studies focusing on $\mu$-variation it is assumed that all atomic masses scale as the proton, \emph{i.e} protons and neutrons treated equally, hence $\mu_{red}$ can be replaced by $\mu$ in this analysis~\cite{Ubachs2007,Reinhold2006}. Hence the derivatives for the level energies can be analytically taken with
\begin{equation}
 \frac{dY_{k,l}}{d\mu} \approx - \frac{Y_{k,l}}{\mu} \left(l + \frac{k}{2} \right)
 \label{Eq-Y-derivative}
\end{equation}
Substitution in Eqs.~(\ref{Eq-energy}) and (\ref{eq-K}) then straightforwardly yields the sensitivity coefficients $K_{\mu}$ for the A-X lines of CO.

Interactions of the A$^1\Pi$ state with triplet states in the CO molecule perturb the level structure~\cite{Field1972}. The level shifts play an important role in the comparative analysis of quasar data to extract $\mu$ variation, but these are implicitly included in the experimental determination of spectral line positions, \emph{i.e.} in the values of Table~\ref{Table-CO-lines-results}. However, in the calculation of $K_{\mu}$ coefficients the admixtures of perturbing character into the wave function composition in A$^1\Pi$ states should also be accounted for. In Refs.~\cite{Ubachs2007,Reinhold2006} a model is proposed to calculate the $K_{\mu}$ coefficients in the case of non-adiabatic mixing between B$^1\Sigma_u^+$ and C$^1\Pi_u$ levels in H$_2$. This model can be adopted for CO and in good approximation one may derive coefficients
\begin{equation}
 K_{\mu} = \alpha_{pure} K_{pure} +  \sum \alpha_{pert} K_{pert}
 \label{K-mix}
\end{equation}
where $K_{pure}$ refers to $K_{\mu}$ coefficients for transitions of the state in consideration and $K_{pert}$ refer to those of the perturber states, while $\alpha$ refers to the admixture in the wave function composition (given in Ref.~\cite{Morton1994}).

\begingroup
\squeezetable
\begin{table*}
\caption{$K_{\mu}$ sensitivity coefficients for $^{12}$C$^{16}$O $A^1\Pi$ - $X^1\Sigma^+$ ($v'-v''$) bands. The uncertainty is estimated to be better than $1\%$.
\label{Klist}}
\begin{ruledtabular}
\begin{tabular}{crrrrrrrrr}
$J''$	&\multicolumn{1}{c}{R} &\multicolumn{1}{c}{Q} &\multicolumn{1}{c}{P} %
		&\multicolumn{1}{c}{R} &\multicolumn{1}{c}{Q} &\multicolumn{1}{c}{P} %
		&\multicolumn{1}{c}{R} &\multicolumn{1}{c}{Q} &\multicolumn{1}{c}{P} \\
\colrule
\\
 	&\multicolumn{3}{c}{\textbf{(0-0)}} 	   &	\multicolumn{3}{c}{\textbf{(1-0)}}		&    \multicolumn{3}{c}{\textbf{(2-0)}}\\		                      	 
0	&-0.00232	&		&	   &	0.01312		&		&		&    0.01850	&	        &       \\
1	&-0.00227	&-0.00237	&          &    0.01280    	&0.01306	&		&    0.01853	&0.01844	&       \\
2	&-0.00223	&-0.00238	&-0.00249  &    0.01235    	&0.01269	&0.01294	&    0.01855	&0.01842	&0.01833\\
3	&-0.00219	&-0.00239	&-0.00257  &    0.01183    	&0.01218	&0.01251	&    0.01856	&0.01838	&0.01825\\
4	&-0.00216	&-0.00240	&-0.00264  &    0.01129    	&0.01160	&0.01195	&    0.01856	&0.01834	&0.01816\\
5	&-0.00212	&-0.00239	&-0.00273  &    0.01077    	&0.01100	&0.01132	&    0.01854	&0.01828	&0.01806\\
6	&-0.00207	&-0.00234	&-0.00281  &    0.01029    	&0.01042	&0.01066	&    0.01847	&0.01821	&0.01795\\
7	&-0.00199	&-0.00219	&-0.00289  &    0.00989    	&0.00989	&0.01002	&    0.01847	&0.01813	&0.01782\\
8	&-0.00185	&-0.00168	&-0.00296  &    0.00954    	&0.00943	&0.00943	&    0.01843	&0.01804	&0.01763\\
\\
	&\multicolumn{3}{c}{\textbf{(3-0)}}	   &	\multicolumn{3}{c}{\textbf{(4-0)}}		&    \multicolumn{3}{c}{\textbf{(5-0)}}\\
0	&0.02756	&		&	   &	0.03784		&		&		&    0.04329	&		&	\\
1	&0.02759	&0.02750	&          &    0.03783	   	&0.03697	&               &    0.04331	&0.04324	&       \\
2	&0.02761	&0.02748	&0.02740   &    0.03771	   	&0.03680	&0.03695        &    0.04332	&0.04321	&0.04314\\
3	&0.02761	&0.02744	&0.02732   &    0.03749	   	&0.03659	&0.03685        &    0.04332	&0.04317	&0.04306\\
4	&0.02761	&0.02739	&0.02723   &    0.03720	   	&0.03636	&0.03667        &    0.04331	&0.04312	&0.04297\\
5	&0.02759	&0.02734	&0.02712   &    0.03690	   	&0.03614	&0.03643        &    0.04328	&0.04305	&0.04286\\
6	&0.02756	&0.02726	&0.02701   &    0.03661	   	&0.03594	&0.03614        &    0.04324	&0.04297	&0.04274\\
7	&0.02751	&0.02718	&0.02688   &    0.03636	   	&0.03575	&0.03583        &    0.04318	&0.04288	&0.04261\\
8	&0.02746	&0.02708	&0.02675   &    0.03614	   	&0.03556	&0.03552        &    0.04311	&0.04277	&0.04247\\
\\
	&\multicolumn{3}{c}{\textbf{(6-0)}} 	&	\multicolumn{3}{c}{\textbf{(7-0)}}		&    \multicolumn{3}{c}{\textbf{(8-0)}}\\
0	&0.05039	&		&		&0.05612	&		&		&0.06162	&		&		\\
1	&0.05044	&0.05034	&               &0.05614	&0.05607	&               &0.06163	&0.06157	&               \\
2	&0.05051	&0.05034	&0.05024        &0.05615	&0.05605	&0.05598        &0.06169	&0.06154	&0.06147       \\
3	&0.05061	&0.05034	&0.05019        &0.05614	&0.05600	&0.05590        &0.06162	&0.06149	&0.06139       \\
4	&0.05066	&0.05037	&0.05016        &0.05611	&0.05594	&0.05581        &0.06159	&0.06143	&0.06136        \\
5	&0.05124	&0.05048	&0.05016        &0.05607	&0.05587	&0.05570        &0.06154	&0.06135	&0.06119       \\
6	&0.05481	&0.05097	&0.05011        &0.05602	&0.05578	&0.05558        &0.06148	&0.06120	&0.06106       \\
7	&0.05435	&0.05452	&0.05059        &0.05595	&0.05568	&0.05544        &0.06135	&0.06109	&0.06092       \\
8	&0.05202	&0.05381	&0.05407        &0.05587	&0.05556	&0.05529        &0.06132	&0.06103	&0.06077       \\
\\
	&\multicolumn{3}{c}{\textbf{(9-0)}} 	&	\multicolumn{3}{c}{\textbf{(10-0)}}	 							\\
0	&0.06646	&		&		&0.07078	&		&								\\
1	&0.06647	&0.06641	&               &0.07079	&0.07074	&                                                               \\
2	&0.06647	&0.06638	&0.06632        &0.07079	&0.07070	&0.07064                                                       \\
3	&0.06646	&0.06633	&0.06624        &0.07077	&0.07065	&0.07057                                                       \\
4	&0.06642	&0.06627	&0.06615        &0.07073	&0.07059	&0.07047                                                       \\
5	&0.06638	&0.06619	&0.06604        &0.07068	&0.07051	&0.07036                                                       \\
6	&0.06631	&0.06610	&0.06591        &0.07061	&0.07041	&0.07023                                                       \\
7	&0.06623	&0.06599	&0.06577        &0.07053	&0.07029	&0.07009                                                       \\
8	&0.06614	&0.06586	&0.06561        &0.07042	&0.07016	&0.06993                                                       \\
\end{tabular}
\end{ruledtabular}
\end{table*}
\endgroup

Perturbation-corrected $K_{\mu}$ coefficients were calculated using Dunham representations for the perturbing d$^3\Delta$ state from Ref.~\cite{Herzberg1970}, and e$^3\Sigma^-$, I$^1\Sigma^-$, $\textrm{a}'^3\Sigma^+$ states from Ref.~\cite{Simmons1971}, (we have collected the Dunham coefficients in the Supplementary material~\cite{Supp}).  The second-order mass-dependent effect (adiabatic correction) is estimated to be at the $5\times10^{-5}$ level and has been neglected.

\begin{figure}
\resizebox{0.48\textwidth}{!}{\includegraphics{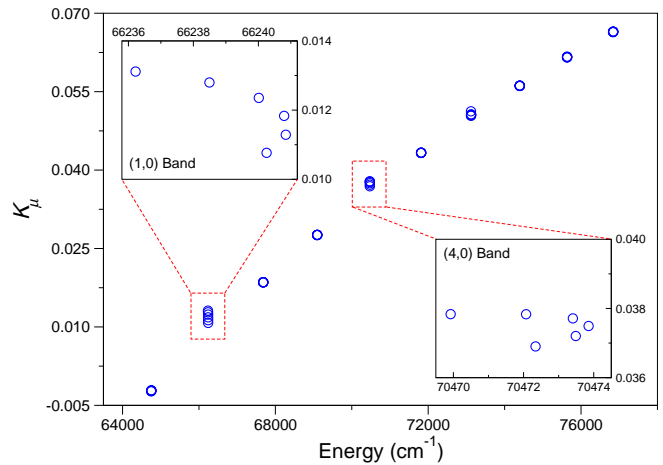}}
\caption{Calculated $K_{\mu}$ coefficients for the spectral lines in the A-X system of CO. The spread in $K_{\mu}$ values, shown enlarged in the insets for the (1,0) and (4,0) bands, illustrate the effect of perturber states.}
\label{K-CO}
\end{figure}

As follows from Eq.~(\ref{K-mix}) the correction to the $K_{\mu}$ coefficients depends on both the amount of admixture of the perturbing state in the wave function, as well as on the value of the $K_{\mu}$ coefficients for the pure perturber state. The most pronounced perturbation occurs between A$^1\Pi$ ($v=1$) and d$^3\Delta$ ($v=5$) with the highest triplet admixture of $17\%$ for the $J'=1$ state. In addition, the perturber $K_{\mu}$ value is more than four times that of the pure A state value, resulting in a correction of $55\%$ to the $K_{\mu}$ value compared to the case when the perturbation is not included. The A$^1\Pi$ ($v=0$) state is only very slightly perturbed by the d$^3\Delta$ ($v=3,4$) states with $0.13\%$ wave function admixture. However, the perturber $K_{\mu}$ is ~10 times larger than that of the pure A state value, resulting in $1.5\%$ correction for $K_{\mu}$. The A$^1\Pi$ ($v=4$) rotational states can have $K_{\mu}$-corrections up to $5\%$, while the A$^1\Pi$ ($v=6, J=8$) rotational state correction is almost $10\%$. The rest of the bands have $K_{\mu}$-corrections that are less than $1\%$. The $K_{\mu}$ coefficients are represented graphically in Fig.~\ref{K-CO}, where the scatter in the ($v=1,4$ and $6$) bands show the effect of perturbations. The resulting $K_{\mu}$ coefficients are listed in Table~\ref{Klist}. The accuracies in these $K_{\mu}$ coefficients are estimated to be better than $1\%$, with the dominant uncertainty contributions from the perturbations (non-adiabatic corrections) which depends on the quantum numbers $v',J'$ of the upper state.

\section{Discussions\label{sec:Discussions}}

$K_{\mu}$ coefficients for CO A-X are thus found in the range of values -0.002 to +0.066 giving a spread equal to that of the Lyman and Werner bands of H$_2$. This makes the A-X system of CO a good search ground for putting constraints on a possible variation of $\mu$, in particular at look-back times of 9.5-11.5 billion years where these features have been detected so far~\cite{Srianand2008,Noterdaeme2009,Noterdaeme2010,Noterdaeme2011}. A set of zero-rest frame wavelengths of A-X ($v=0-9,0$) bands has been measured with most transitions at an accuracy of $\Delta\lambda/\lambda = 1.5 \times 10^{-7}$, and for the (0,0) and (1,0) bands even as accurate as $3 \times 10^{-8}$. The combined data sets of rest-frame wavelengths and sensitivity coefficients are the basis for a $\mu$-variation analysis on CO spectra from high-redshift galaxies, where the analysis will be solely constrained by the quality of the astronomical data. So far six high redshift galaxies have been observed with characteristic CO A-X features. The best example is that of a CO spectrum observed at $z=2.69$ with a column density of $\log N$(CO)$=14.17$ cm$^{-2}$ in the sight-line of the Q1237+064 quasar system, showing a signal-to-noise ratio of 10-40 (over the wavelength range) and resolution $R \sim 50000$ after 8 hrs observation at the ESO Very Large Telescope~\cite{Noterdaeme2010}. Extended observations with attached Th-Ar calibration of the astronomical exposures (not available yet for the discovery spectrum of Ref.~\cite{Noterdaeme2010}) would result in a competitive result on $\Delta\mu/\mu$ from the CO spectra, \emph{i.e.} an estimated constraint at $\Delta\mu/\mu < 10^{-5}$. The objects where CO is detected exhibit large column densities of H$_2$ and HD molecules, since extragalactic abundance ratios are generally $N$(H$_2$)/$N$(CO) $> 10^4$ and $N$(HD)/$N$(CO) $\sim 1$. Hence, future comprehensive $\mu$ constraining analyses can be performed from the features of all three molecules contained in the same quasar spectrum. In view of the increased number of spectral lines, the wider wavelength coverage, and the CO lines lying outside the Lyman-$\alpha$ forest, well-calibrated (Th-Ar attached) and good signal-to-noise observations (SNR ~ 40) should lead to constraints at $\Delta\mu/\mu < 3 \times 10^{-6}$. The system Q1237+064 would be the first target of choice.

\section{Conclusions\label{sec:Conclusions}}

We have identified the $\textrm{A}^1\Pi-\textrm{X}^1\Sigma^+$ band system of carbon monoxide a novel probe system to search for possible variations of the proton-electron mass ratio ($\mu$) on cosmological time scales.
Laboratory wavelengths of the spectral lines of the A-X ($v$,0) bands for $v=0-9$ have been determined at an accuracy of $\Delta\lambda/\lambda=1.5 \times 10^{-7}$ through VUV Fourier-transform absorption spectroscopy, providing a comprehensive and accurate zero-redshift data set.
Two-photon Doppler-free laser spectroscopy has been applied for the (0,0) and (1,0) bands, achieving $3 \times 10^{-8}$ accuracy level. Accurate sensitivity coefficients $K_{\mu}$ for a varying $\mu$ have been calculated for the CO A-X bands, where the effect of perturbations have been accounted for.
It is expected that future $\mu$ constraining analyses that include H$_2$ and HD and CO transitions from the same absorber should result in a more accurate and robust constraint for $\Delta\mu/\mu$.

\smallskip
\begin{acknowledgments}
This work was supported by the Netherlands Astrochemistry Program of NWO (CW-EW).
We are indebted to the general staff of the SOLEIL synchrotron for smoothly running the facility.
We thank Dr. R.W. Field for fruitful discussions.
\end{acknowledgments}


\begin{thebibliography}{}

\bibitem{Malec2010} A.L. Malec, R. Buning, M.T. Murphy, N. Milutinovic, S.L. Ellison,
J.X. Prochaska, L. Kaper, J. Tumlinson, R.F. Carswell, and W. Ubachs,
 MNRAS {\bf 403}, 1541 (2010).%
\bibitem{Weerdenburg2011} F. van Weerdenburg, M.T. Murphy, A.L. Malec, L. Kaper, and W. Ubachs,
 \prl {\bf 106}, 180802 (2011).
\bibitem{Philip2007} J. Philip, J.P. Sprengers, Th. Pielage, C.A. de Lange, W. Ubachs, and E. Reinhold,
  Can. J. Chem. {\bf 82}, 713 (2004).
\bibitem{Salumbides2008} E.J. Salumbides, D. Bailly, A. Khramov, A.L. Wolf, K.S.E. Eikema, M. Vervloet, and W. Ubachs,
 \prl {\bf 101}, 223001 (2008).
\bibitem{Ivanov2008} T.I. Ivanov, M. Roudjane, M.O. Vieitez, C.A. de Lange, W.U.L. Tchang-Brillet, and W. Ubachs,
  \prl {\bf 100}, 093007 (2008).
\bibitem{Varshalovich1993} D.A. Varshalovich and S.A. Levshakov,
   JETP Lett. {\bf 58}, 237 (1993).
\bibitem{Meshkov2006} V. V. Meshkov, A. V. Stolyarov, A. Ivanchik, and D. A. Varshalovich,
   JETP Lett. {\bf 83}, 303 (2006).
\bibitem{Ubachs2007} W. Ubachs, R. Buning, K.S.E. Eikema, and E. Reinhold,
 \jms {\bf 241}, 155 (2007).
\bibitem{Ivanov2010} T.I. Ivanov, G.D. Dickenson, M. Roudjane, N. de Oliveira, D. Joyeux, L. Nahon, W.U.L. Tchang-Brillet,
 and W. Ubachs, Mol. Phys. {\bf 108}, 771 (2010).
\bibitem{Murphy2008} M. T. Murphy, V.V. Flambaum, S. Muller, and C. Henkel,
 Science {\bf 320}, 1611 (2008).
\bibitem{Kanekar2011} N. Kanekar, Astroph. J. Lett. {\bf 728}, L12 (2011).
\bibitem{Henkel2009} C. Henkel, K.M. Menten, M.T. Murphy, N. Jethava, V.V. Flambaum, J.A. Braatz,
  S. Muller, J. Ott, and R.Q. Mao, Astron. Astroph. {\bf 500}, 725 (2009).
\bibitem{Jansen2011} P. Jansen, L.-H. Xu, I. Kleiner, W. Ubachs, and H.L. Bethlem, \prl {\bf 106}, 100801 (2011).
\bibitem{Levshakov2011} S.A. Levshakov, M.G. Kozlov, and D. Reimers, Astroph. J. {\bf 738}, 26 (2012).
\bibitem{Jansen2011b} P. Jansen, I. Kleiner, L.-H. Xu, W. Ubachs, and H.L. Bethlem, \pra {\bf 84}, 062505 (2011).
\bibitem{Muller2011} S. Muller, A. Beelen, M. Gu\'{e}lin, S. Aalto, J.H. Black, F. Combes, S.J. Curran,
  P. Theule, S.N. Longmore, Astron. Astroph. {\bf 535}, A103 (2011).
\bibitem{Ellingsen2012} S.P. Ellingsen, M.A. Voronkov, S.L. Breen, and J.E.J. Lovell,
  Astroph. J. Lett. {\bf 747}, L7 (2012).
\bibitem{Bagdonaite2012} J. Bagdonaite, P. Jansen, C. Henkel, H.L. Bethlem, K.M. Menten, and W. Ubachs, submitted (2012).
\bibitem{Srianand2008} R. Srianand, P. Noterdaeme, C. Ledoux, and P. Petitjean,
 \aa {\bf 482}, L39 (2008).
\bibitem{Noterdaeme2009} P. Noterdaeme, C. Ledoux, R. Srianand, P. Petitjean, and S. Lopez,
 \aa {\bf 503}, 765 (2009).
\bibitem{Noterdaeme2010} P. Noterdaeme, P. Petitjean, C. Ledoux, S. Lopez, R. Srianand, and S.D. Vergani,
 \aa {\bf 523}, A80 (2010).
\bibitem{Noterdaeme2011} P. Noterdaeme, P. Petitjean, R. Srianand, C. Ledoux, and S. Lopez,
 \aa {\bf 526}, L7 (2011).
\bibitem{Prochaska2009} J.X. Prochaska, Y. Sheffer, D.A. Perley, J.S. Bloom, L.A. Lopez, M. Dessauges-Zavadsky, H.-W. Chen,
 A.V. Filippenko, M. Ganeshalingam, W. Li, A.A. Miller, and D. Starr,
 \apjl {\bf 691}, L27 (2009).
\bibitem{Field1972} R.W. Field, S.G. Tilford, R.A. Howard, and J.D. Simmons,
 \jms {\bf 44}, 347 (1972); R.W. Field, B.G. Wicke, J.D. Simmons, and S.G. Tilford,
 \jms {\bf 44}, 383 (1972).
\bibitem{Lefloch1987} A.C. Le Floch, F. Launay, J. Rostas, R.W. Field, C.M. Brown, and K. Yoshino,
 \jms {\bf 121}, 337 (1987);
 A.C. Le Floch, PhD Thesis, Universit\'{e} de Paris-Sud (1992);
 A.C. Le Floch, \jms {\bf 155}, 177 (1992);
 C. Kittrell, A.C. Le Floch, and B.A. Garetz, \jpc {\bf 97}, 2221 (1993).
\bibitem{Morton1994} D.C. Morton and L. Noreau, \apjss {\bf 95}, 301 (1994).
\bibitem{Drabbels1997} M. Drabbels, J.J. ter Meulen and W.L. Meerts, \cpl {\bf 267}, 127 (1997).
\bibitem{Nahon2012} L. Nahon, N. de Oliveira, G. Garcia, J.F. Gil, B. Pilette, O. Marcouille,
 B. Lagarde, and F. Polack, J. Synchrotron Rad. {\bf 19}, 508 (2012).
\bibitem{Oliveira2011} N. de Oliveira, M. Roudjane, D. Joyeux, D. Phalippou, J.C. Rodier, and L. Nahon,
 Nat. Photon. {\bf 5}, 149 (2011).
\bibitem{Saloman2004} E.B. Saloman, J. Phys. Chem. Ref. Data {\bf 33}, 765 (2004).
\bibitem{Humphreys1970} C.J. Humphreys and E. Paul, \josa {\bf 60}, 1302 (1970).
\bibitem{Brandi2001} F. Brandi, I. Velchev, W. Hogervorst, and W. Ubachs,
 \pra {\bf 64}, 032505 (2001).
\bibitem{Ubachs1997} W. Ubachs, K.S.E. Eikema, W. Hogervorst, and P.C. Cacciani,
 \josab {\bf 14}, 2469 (1997).
\bibitem{Xu2000} S. Xu, R. van Dierendonck, W. Hogervorst, and W. Ubachs,
  \jms 201, 256 (2000).
\bibitem{Bodermann2002} We used the "IodineSpec" program, kindly provided to us by Dr. H. Kn\"{o}ckel (Leibniz University, Hannover). See also B. Bodermann, H. Kn\"{o}ckel and E. Tiemann, Eur. Phys. J. D 19, 31, (2002).
\bibitem{Hannemann2006} S. Hannemann, E.J. Salumbides, S. Witte, R.T. Zinkstok, E.-J. van Duijn, K.S.E. Eikema, and W. Ubachs,
 \pra {\bf 74}, 062514 (2006).
\bibitem{Varberg1992} T.D. Varberg and K.M. Evenson, \apj {\bf 385}, 763 (1992).
\bibitem{Guelachvili1983} G. Guelachvili, D. de Villeneuve, R. Farrenq, W. Urban, and J. Verges,
 \jms {\bf 98}, 64 (1983).
\bibitem{Simmons1969} J.D. Simmons, A.M. Bass and S.G. Tilford, \apj {\bf 155}, 345 (1969).
\bibitem{Reinhold2006} E. Reinhold, R. Buning, U. Hollenstein, A. Ivanchik, P. Petitjean, and W. Ubachs,
 \prl {\bf 96}, 151101 (2006).
\bibitem{Herzberg1970} G. Herzberg, T.J. Hugo, S.G. Tilford, and J.D. Simmons,
 \cjp {\bf 48}, 3004 (1970).
\bibitem{Simmons1971} J.D. Simmons and S.G. Tilford, J. Res. Nat. Bur. Stand. Sect. A {\bf 75}, 455 (1971).
\bibitem{Supp} See Supplementary Material at [URL to be inserted by publisher] for further details.

\end{thebibliography}
\end{document}


\title{{{\emph{Supplementary Information for}}} \\[0.2ex] The CO A-X Absorption Bands for Constraining Cosmological Drift \\ of the Proton-Electron Mass Ratio}

\author{E. J. Salumbides}
\affiliation{Department of Physics and Astronomy, and LaserLaB, VU University, De Boelelaan 1081, 1081 HV Amsterdam, The Netherlands}
\affiliation{Department of Physics, University of San Carlos, Cebu City 6000, Philippines}
\author{M. L. Niu}
\affiliation{Department of Physics and Astronomy, and LaserLaB, VU University, De Boelelaan 1081, 1081 HV Amsterdam, The Netherlands}
\author{J. Bagdonaite}
\affiliation{Department of Physics and Astronomy, and LaserLaB, VU University, De Boelelaan 1081, 1081 HV Amsterdam, The Netherlands}
\author{N. de Oliveira}
\affiliation{Synchrotron Soleil, Orme des Merisiers, St Aubin BP 48, 91192, GIF sur Yvette cedex, France}
\author{D. Joyeux}
\affiliation{Synchrotron Soleil, Orme des Merisiers, St Aubin BP 48, 91192, GIF sur Yvette cedex, France}
\author{L. Nahon}
\affiliation{Synchrotron Soleil, Orme des Merisiers, St Aubin BP 48, 91192, GIF sur Yvette cedex, France}
\author{W. Ubachs}
\affiliation{Department of Physics and Astronomy, and LaserLaB, VU University, De Boelelaan 1081, 1081 HV Amsterdam, The Netherlands}

\maketitle

\begin{table}
   \centering
    \caption{Dunham coefficients $Y_{kl}$ for the states of $^{12}$C$^{16}$O involved in the present analysis: the X$^1\Sigma^+$ ground state, the A$^1\Pi$ excited state and the perturber states. The constants were obtained from~\cite{Guelachvili1983},~\cite{Simmons1969},~\cite{Herzberg1970}, and~\cite{Simmons1971} respectively. All values in cm$^{-1}$.}
    \begin{tabular}{l@{\hspace{10pt}}l@{\hspace{15pt}}l@{\hspace{15pt}}l@{\hspace{15pt}}l@{\hspace{15pt}}l@{\hspace{15pt}}l}
    \hline
\multicolumn{1}{c}{$Y_{k\,l}$} & \multicolumn{1}{c}{X$^1\Sigma^+$}     & \multicolumn{1}{c}{A$^1\Pi$}  	& \multicolumn{1}{c}{d$^3\Delta$}   & \multicolumn{1}{c}{e$^3\Sigma^-$}	& \multicolumn{1}{c}{a'$^3\Sigma^+$}	& \multicolumn{1}{c}{I$^1\Sigma^-$}	\bigstrut \\
    \hline
$Y_{1\,0}$ &  0.2169813079E+04 &  0.1518240E+04	&  0.116818E+04	&  0.111772E+04	&  0.12286E+04	& 0.109222E+04	\\
$Y_{2\,0}$ & -0.1328790597E+02 & -0.194E+02	& -0.970E+01	& -0.10686E+02	& -0.10468E+02	& -0.10704E+02	\\
$Y_{3\,0}$ &  0.1041444739E-01 &  0.76584E-00  	&		& 		& 		& 		\\
$Y_{4\,0}$ &  0.6921598529E-04 & -0.14117E-00 	&		& 		& 		& 		\\
$Y_{5\,0}$ &  0.1657890319E-06 &  0.1434E-01  	&		& 		& 		& 		\\
$Y_{6\,0}$ &  0.2466226718E-08 & -0.8051E-03 	&		& 		& 		& 		\\
$Y_{7\,0}$ & -0.8630071431E-09 &  0.236E-04   	&		& 		& 		& 		\\
$Y_{8\,0}$ &  0.1261536024E-10 & -0.29E-06   	&		& 		& 		& 		\\
$Y_{9\,0}$ & -0.8363842345E-13 &  		&		& 		& 		& 		\\
$Y_{0\,1}$ &  0.1931280862E+01 &  0.16115E+01  	&  0.13089E+01	&  0.12836E+01	&  0.13446E+01	& 0.12705E+01	\\
$Y_{1\,1}$ & -0.1750410155E-01 & -0.23251E-01 	& -0.1668E-01	& -0.1753E-01	& -0.1892E-01	& -0.1848E-01	\\
$Y_{2\,1}$ &  0.5422101371E-06 &  0.15911E-02   &		& 		& 		& 		\\
$Y_{3\,1}$ &  0.1311844382E-07 & -0.57160E-03  	&		& 		& 		& 		\\
$Y_{4\,1}$ &  0.1401093763E-08 &  0.82417E-04   &		& 		& 		& 		\\
$Y_{5\,1}$ & -0.5329907475E-11 & -0.59413E-05  	&		& 		& 		& 		\\
$Y_{6\,1}$ & -0.1434127145E-11 &  0.21149E-06   &		& 		& 		& 		\\
$Y_{7\,1}$ & 		       & -0.2991E-08   	&		& 		& 		& 		\\
$Y_{0\,2}$ & -0.6120747566E-05 & 		& -0.60E-05	& -0.677E-05	& -0.641E-05	& -0.90E-5	\\
$Y_{1\,2}$ &  0.9449843095E-09 &	\\
$Y_{2\,2}$ & -0.1430768382E-09 &	\\
$Y_{3\,2}$ & -0.2927592559E-11 &	\\
$Y_{4\,2}$ &  0.1660533203E-12 &	\\
$Y_{5\,2}$ & -0.4714582132E-14 &	\\
$Y_{0\,3}$ &  0.5555386989E-11 &	\\
$Y_{1\,3}$ & -0.1512463732E-12 &	\\
$Y_{2\,3}$ & -0.1471295100E-14 &	\\
\hline
\hline
    \end{tabular}
    \label{dunham}
\end{table}